# METHODS OF USING EXISTING AND CURRENTLY USED WIRE LINES (POWER LINES, PHONE LINES, INTERNET LINES) FOR TOTALLY SECURE CLASSICAL COMMUNICATION UTILIZING KIRCHOFF'S LAW AND JOHNSON-LIKE NOISE[1]

LASZLO B. KISH[+]

[+]*Department of Electrical Engineering, Texas A&M University, College Station, TX 77843-3128, USA*

We outline some general solutions to use already existing and *currently used* wire lines, such as power lines, phone lines, internet lines, etc, for the unconditionally secure communication method based on Kirchoff's Law and Johnson-like Noise (KLJN). Two different methods are shown. One is based on filters used at single wires and the other one utilizes a common mode voltage superimposed on a three-phase powerline.

## 1. Introduction

Recently, a totally secure classical communication scheme was introduced [1,2] utilizing two identical pairs of resistors and noise voltage generators, the physical properties of an idealized Kirchoff-loop and the statistical physical properties thermal noise (KLJN communicator). The resistors (low bit = small resistor $R_L$, high bit = large resistor $R_H$) and their thermal-noise-like voltage generators (thermal noise voltage enhanced by a pre-agreed factor) are randomly chosen and connected at each clock period at the two sides of the wire channel. A *secure bit exchange* takes place when the states at the two ends are different, which is indicated by an intermediate level of the *rms* noise voltage on the line, or that of the *rms* current noise in the wire. The most attractive properties of the KLJN cipher are related to its *security* [1, 3-7] and the *robustness* of classical information. In the *idealized* scheme of the Kirchoff-loop-Johnson(-like)-noise cipher, the passively observing eavesdropper can extract *zero bit* of information and the actively eavesdropping observer can extract at most *one bit* before getting discovered [3-7]. Eavesdropper executing a man-in-the-middle attack can extract virtually zero bit before getting discovered [3]. The KLJN system has recently became network-ready [6]. This new property [6] opens a large scale of practical applications because the KLJN cipher can be installed as a computer card [6], similarly to Eternet cards.

Quantum communication, the alternative communication line secured by physical laws, needs a separate, well isolated optical cable (so-called "dark optical fiber") to function because of the fragility of quantum information based on single photon operations. Similarly, up to now, it has been a common assumption that the KLJN communication also needs separate wire lines to operate because the Kirchoff loops of communicator pars must stay single loops as given in the original description in the KLJN scheme.

However, via a few examples we will show that existing and used wire lines, such as power lines, phone wire lines and internet wire lines can also be used to build a KLJN network by implementing proper filter units at each intersection within the line section of concern. We avoid discussing any details of the filters here; we suppose they are idealized filters passing only the allowed bands of frequencies. We suppose that the frequency ranges of the KLJN frequency band and that of the normal usage of the wire are established and they do not overlap. For power lines that means that for large distances the frequency range of the KLJN cipher must be below 60 Hz (or 50 Hz).

---

[1] A TAMU patent disclosure is submitted.

## 2. The filter method for single lines

In Figure 1, an example is shown how to use KLJN frequency Band Excluder (BE) and Band Pass (BP) filters to preserve a single Kirchhoff's loop characteristics in the KLJN frequency band between two KLJN communicators with one intersection between them. Both, the original non-KLJN load (power consumer, phone, or internet card represented by $R_N$) and the KLJN communicators will work using their own, non-overlapping frequency bands. BE excludes the KLJN frequency band and passes all other important frequencies. BP passes the KLJN frequency band and excludes all other frequencies. BP filters must be used in any case at the communicator outputs to avoid transient and eavesdropping probing signals out of the KLJN band.

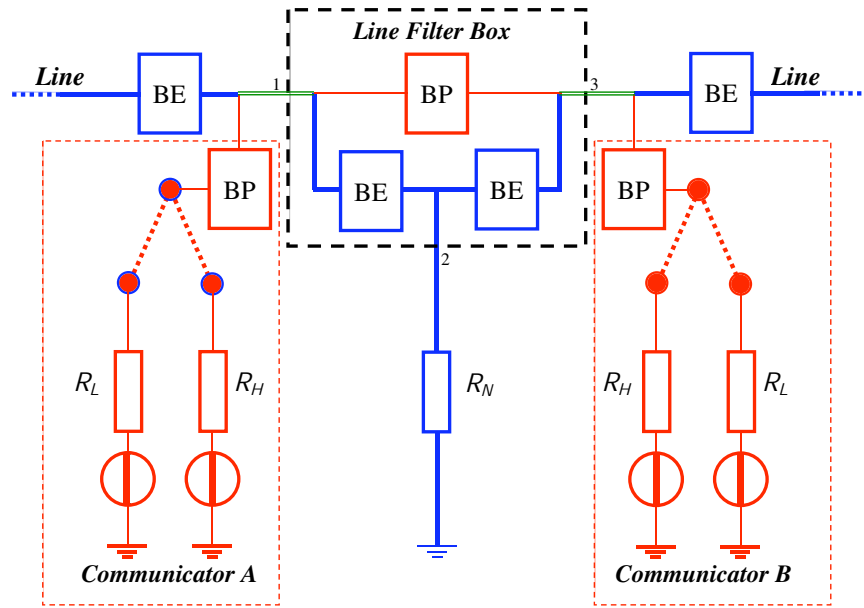

**Figure 1.** Example for how to use KLJN frequency Band Excluder (BE) and Band Pass (BP) filters to preserve a single Kirchhoff loop in the KLJN frequency band between two KLJN communicators with one intersection between them. Thick (blue) lines: original line current; thin (red) line: KLJN current; double (green) lines: both types of currents.

Figure 2 shows that, though the topology in Figure 1 may look complex, the circuitry in the Line Filter Box in Figure 1 can be contained by a box with only three electrodes. if there are more then one consumers along the line between the communicators, each of them must be fed via a separate Line Filter Box.

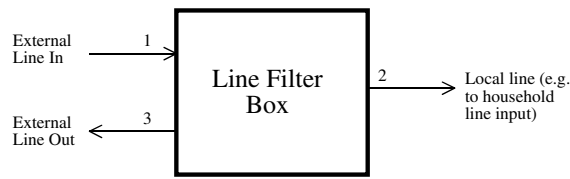

**Figure 2.** The line filter box (see Figure 1) should be installed at each intersection of the line to separate the non-KLJN communicator loads from the KLJN frequency band.

## 3. Common voltage method with 3-phase power lines

In the case of 3-phase power lines, we can make use of the fact that, in the idealized case, when the load is symmetric on the phases, there is no current flowing from the common point of the 3-phase transformers to the ground. Then the KLJN communicators can be connected between the common points of the 3-phase transformers and the ground, see Figure 3. This arrangement can help to reduce the problem of working with high voltages because the common point of the transformers is at ground potential at the idealized case.

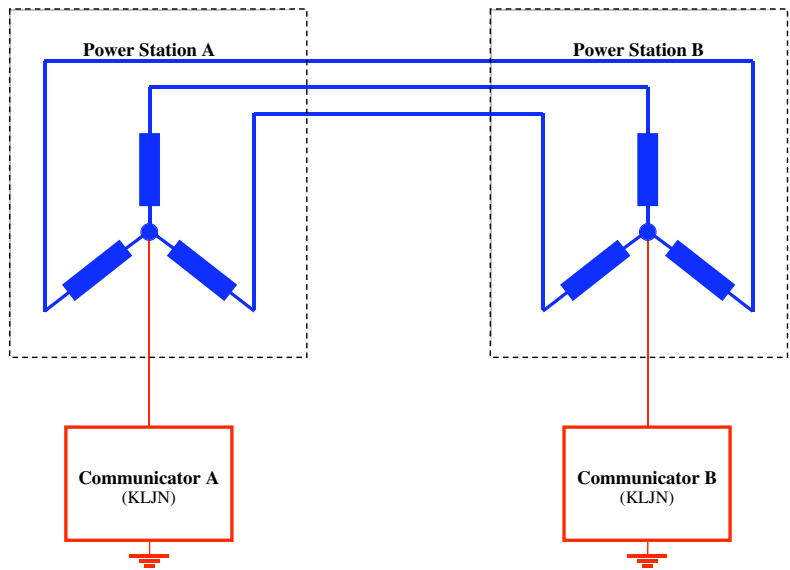

**Figure 3.** Communication via idealized 3-phase power lines with symmetric loads of the 3-phase transformers at Power Stations A and B, respectively.

However, in the practice, there are non-idealities and asymmetries among the lines (load, phase, etc.) therefore filters will be necessaries to avoid problems. Figure 4 shows a possible solution of such problems. In the KLJN frequency range the filters drive the current through the communicators and out of that frequency range the current goes into the ground. Note, if there is an intersection of cables with asymmetric loads between Power Stations A and B then filters must be used similarly to Figure 1 and then the problems with high voltages cannot be avoided. Therefore, the arrangements in Figures 3 and 4 are most advantageous when there is no intersection between the two power stations.

## Summary

We have shown with some simple circuit demonstrations that the KLJN communicators can use existing and currently used wires for communication. We presented some examples as demonstration. The filter method can be used also with phone and internet lines and getting the KLJN current around switchers and multiplexer units. The concrete realization and further developments of these ideas is straightforward, though by no means it is trivial, and they reach out of the scope of the present paper.

Finally, we must answer the following questions. Is the communication still secure if the eavesdropper removes a filter or if a non-filtered new intersection is made on the line? The answer is straightforward: the communication becomes non-secure however the current-voltage alarm system [3], which is comparing the KLJN voltages and currents at the two communicators, will go on and the communication will immediately be terminated for security reasons therefore the eavesdropper can at most extract a single bit of information [3].

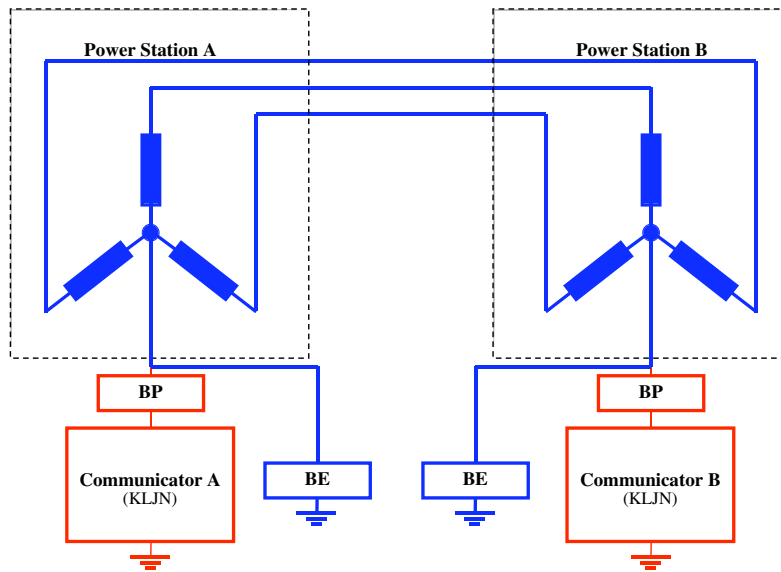

**Figure 3.** Communication via practical 3-phase power lines.

## Acknowledgment

A discussion with Gabor Schmera is appreciated.